\documentclass[aps,pra,english,reprint,longbibliography,nofootinbib,floatfix]{revtex4-2}
\usepackage[utf8]{inputenc}
\usepackage[T1]{fontenc}
\usepackage[stretch=20]{microtype}
\usepackage{graphicx}
\usepackage[shortlabels]{enumitem}
\usepackage[draft=false, hidelinks]{hyperref}

\usepackage{nccmath}
\usepackage{mathtools} % load after nccmath
\usepackage{isomath}
\usepackage{relsize}
\usepackage{scalerel}
\usepackage{amssymb}
\usepackage{mathrsfs}
\usepackage{siunitx}
\usepackage{braket}
\usepackage{mleftright}
\mleftright

\renewcommand*\vec[1]{\mathbfit{#1}}

\AtBeginDocument{}

\newcommand*\defeq{\vcentcolon=}

\DeclareMathOperator\erf{erf}
\DeclareMathOperator\erfcx{erfcx}
\DeclareMathOperator*\symm{\scaleobj{0.8}{\scalerel*{\rm S}{\sum}}}
\AtBeginDocument{
  \let\Re\relax
  
  \DeclareMathOperator\Re{Re}
  
}

\DeclareMathOperator\tr{tr}
\DeclareMathOperator\rank{rank}
\DeclareMathOperator\detprime{det^\prime}
\newcommand*\fock[1]{\mathscr{#1}}
\newcommand*\VIB{V}
\newcommand*\bound{\mathrm{b}}

\begin{document}

\title{Bosonic Functional Determinant Approach and its Application to Polaron Spectra}

\author{Moritz Drescher}
\author{Manfred Salmhofer}
\author{Tilman Enss}
\affiliation{Institut f\"ur Theoretische Physik,
  Universit\"at Heidelberg, 69120 Heidelberg, Germany}

\begin{abstract}
The functional determinant approach (FDA) is a simple method to compute exactly certain observables
for ideal quantum systems and has been successfully applied to the Fermi polaron problem to
obtain the dynamical overlap and spectral function.
Unfortunately, its application to Bosonic ultracold gases is prohibited by the failure of the
grand canonical ensemble for these systems.
In this paper, we show how to circumvent this problem and develop a Bosonic FDA.
This yields exact injection and ejection spectra for ideal Bose polarons at arbitrary temperatures.
While coherent features visible at absolute zero quickly smear out with rising temperature as expected,
the line width of the main peak is, counterintuitively, found to decrease near unitarity.
Furthermore, we provide explicit formulas for the overlap operator, which allow to compute the necessary
determinants for both Bose and Fermi polarons more efficiently than previously possible.
\end{abstract}

\maketitle

\section{Introduction}

Ideal quantum systems exhibit interesting many-body physics while
remaining simple enough for exact treatment in many aspects.
The Bose and Fermi gas are among the canonical examples.
The insertion of an impurity atom into these systems allows to study their reaction to strongly localized perturbations
and constitutes a paradigmatic instance of the formation of a quasi-particle, a polaron.

Progress in experiments with ultracold gases has allowed to realize polarons in Bose-Einstein condensates
\cite{Rentrop2016, Hu2016, Jorgensen2016, Yan2020, Skou2021, Cayla2023, Etrych2024} and much of their rich behaviour
can be understood already at the level of ideal gas models.
Many theoretical works have studied the Bose polaron problem in the general setting of an interacting gas and mobile impurities
using variational Ansätze; we refer exemplarily to
\cite{Girardeau1961, Tempere2009, Rath2013, Shchadilova2016, Drescher2019, Drescher2020, Levinsen2021, Christianen2022, Rose2022}.
Most of these methods are limited to zero temperature, two particularly noteworthy exceptions being \cite{Dzsotjan2020, Field2020}.
In the case of an ideal gas and immobile impurity, the “ideal Bose polaron”, the knowledge of the single-particle wave functions 
has allowed to obtain exact results at non-zero temperature, but only for single-particle observables like energy or Tan’s contact \cite{Drescher2021}.
For the Fermi polaron problem, on the other hand, even many-particle observables like the dynamical overlap and
spectral function, both of particular experimental relevance, could be computed exactly using a technique called functional determinant approach (FDA)
\cite{Knap2012, Cetina2016, Liu2020}.

The FDA is a simple method which allows to express certain many-body observables of ideal quantum systems in
terms of their single-particle eigenstates.
However, it requires \textit{a priori} the use of the grand canonical ensemble (GCE).
Unfortunately, the GCE is not valid for a Bose gas below critical temperature,
as it predicts macroscopic fluctuations of the condensate density \cite{Ziff1977} –
this is the so-called grand-canonical catastrophe.

In this paper, it is shown how this difficulty can be circumvented and thus how the FDA
can be extended to Bosonic systems.
It is then applied to compute overlaps and spectra of ideal Bose polarons at arbitrary temperatures.
We study in particular how the peak positions and widths depend on temperature
and find that, at strong coupling, the peaks become sharper for hotter systems.
The relevance of correlations between condensate and thermal gas is investigated and found to provide the leading-order correction
near zero and the critical temperature when compared to considering only the dominant component.

Compared to earlier finite-temperature works, our approach is exact but limited to ideal Polarons.
In contrast, \cite{Dzsotjan2020, Field2020} employ variational Ansätze while \cite{Schmidt2016} uses a heuristic
extension of the FDA that neglects correlations between condensate and thermal gas.
\cite{Drescher2021} is also an exact study of the ideal Bose Polaron,
but its finite-temperature results are limited to single-particle observables.

\paragraph*{Outline}
First, the general method is described briefly in Section II and it is shown how to apply it to the
problem at hand.
The resulting Bose polaron spectra are discussed in Section III.
We conclude with a discussion of how the method might be generalized in the future.
Three appendices contain the full derivation of the Bosonic FDA, some general determinant expansion formulas required for it,
and analytical formulas for the matrix elements needed for an efficient computation of the dynamical overlap for Bose and Fermi polarons.

\paragraph*{Terminology note} Throughout the paper, we employ the term \emph{polaron} to refer to an impurity in an
ultracold gas, even though its quasi-particle nature is debated for a Bosonic bath and strong coupling:
Here, the peak in the rf spectrum becomes very broad and the so-called quasi-particle residue vanishes.
On the other hand, it is not clear whether these quantities represent a good measure of being a quasi-particle
when coupling is strong.

\section{Method}

In this section, we describe the method, the Bosonic functional determinant approach.
We start by reviewing the usual FDA, then outline its extension to condensates.
The full derivation can be found in the appendix.

\subsection{Klich's Formula and the FDA}

Klich's formula \cite{Klich2002} is concerned with exponentials of one-particle operators in Fock space.
If $A$ is an operator on single-particle Hilbert space, we denote by
\begin{equation*}
\fock A = \int d^3 \vec{r} \, a_{\vec r}^\dagger (A a)_{\vec r}^{}
\end{equation*}
its corresponding one-particle operator on Fock space where $a_{\vec r}^{(\dagger)}$ are the bosonic or fermionic
creation and annihilation operators.

Klich's formula then states that
\begin{equation}
\tr e^{\fock A} e^{\fock B} = \begin{cases}
		\det (1 + e^A e^B) & \text{ Fermions} \\
		\det (1 - e^A e^B)^{-1} & \text{ Bosons}
	\end{cases}
\end{equation}
and likewise for more than two operators.

In the grand canonical ensemble (GCE), the density operator itself is of the exponential
form with the trace running over the entire Fock space.
This yields the so-called functional determinant approach (FDA):
\begin{align}
\langle e^{\fock A} e^{\fock B} \rangle_\text{gc}
 &= \begin{cases}
	 \det (1 + M n_{F}) & \text{ Fermions} \\
	 \det (1 - M n_{B})^{-1} & \text{ Bosons,}
	\end{cases} \label{eq:FDA}
\end{align}
where
\begin{align*}
	M &= e^A e^B - 1, \\
	n_{F/B} &= \frac{1}{e^{\beta (H - \mu)} \pm 1}.
\end{align*}

This has been successfully applied to the Fermi polaron problem \cite{Knap2012, Cetina2016, Liu2020} to compute exactly the dynamical overlap
\begin{align*}
	S(t) &= \langle e^{it \fock H_i} e^{-it\fock H_f} \rangle,
\end{align*}
where $\fock H_i$ and $\fock H_f$ are initial and final Hamiltonians
of a sudden change (a \emph{quench}) in coupling strength.
From $S$, one could then obtain its Fourier transform, the spectral function $A$.
For Bosons, on the other hand, \eqref{eq:FDA} cannot be applied below the critical temperature
because of the grand-canonical catastrophe:
in the GCE, the condensate density $n_0$ does not assume
a well-defined value, but follows an exponential distribution.
If we tried to compute an expectation value $\braket O$ using the GCE, we would instead obtain its Laplace transform
in $n_0$, $\braket{O}_\text{gc} = \frac{1}{\braket{n_0}_\text{gc}} \int e^{-n_0 / \braket{n_0}_\text{gc}} \braket{O}_{n_0} dn_0$,
which does not correspond to the experimental situation of fixed $n_0$ \cite{Kristensen2019, Christensen2021}.

\subsection{Bosonic FDA}

We have found a replacement for \eqref{eq:FDA} that works below the critical temperature.
Instead of the GCE,
we use an ensemble that is grand canonical for the non-zero modes but has a fixed
number $N_0$ of condensed particles.
Its density operator and partition function are given by
\begin{align*}
	\rho &= \frac{\delta_{\fock N_0, N_0} e^{-\beta \fock H_{gc}^\prime}}{Z},  \\
	Z &= \tr^\prime e^{-\beta \fock H_{gc}^\prime}
		= \detprime (1 - e^{-\beta H_{gc}^\prime})^{-1}.
\end{align*}
where $A^\prime$ denotes the operator $A$ acting only on the space orthogonal to the zero mode,
$\det\nolimits^\prime$
the determinant over this space,
$\fock A'$ the Fock space operator corresponding to $A'$, which thus acts only on the
subspace with zero particles in the zero mode, and
$\tr^\prime$ the trace over this space.
Finally, $H_{gc} = H - \mu$, so $\fock H_{gc} = \fock H - \mu \fock N$.
This and similar ensembles have been used to study the particle number fluctuations of the
Bose gas \cite{Politzer1996, Navez1997, Kruk2023}.

With this, we obtain the following \emph{Bosonic FDA} (derivation in the appendix):
\begin{align}
	\langle e^{\fock A} e^{\fock B} \rangle &= S_0 S_\mathrm{t} S_\mathrm{c} \label{eq:BFDA}
\end{align}
where
\begin{align*}
	S_0 &= \exp \bra{\Phi_0} M \ket{\Phi_0} \\
	S_\mathrm t &= \det (1 - M n_B)^{-1} \\
	S_\mathrm c &= \exp\left(\frac{\det (1 - M n_B + M \ket{\Phi_0} \bra{\Phi_0} M n_B)}{\det (1 - M n_B)} - 1 \right)
\end{align*}
and likewise for more than two operators.
Here, $\Phi_0 = \sqrt{N_0} \phi_0$ is the condensate wave function
where $\phi_0$ is the single-particle ground state, usually constant.
The expressions hold in the thermodynamic limit; in particular $\phi_0$ is not 
included in the determinants because it is not a normalizable vector.%
\footnote{But $\sqrt{N_0} \phi_0$ converges pointwise in the thermodynamic limit and expressions involving $\ket{\Phi_0}$, $\bra{\Phi_0}$
are well-defined as position space integrals.}

The result consists of three factors:
$S_0$ is related to the condensate and the only factor present at zero temperature.
$S_\mathrm t$ captures the effect of the thermally excited modes and is the only term above the critical temperature, where
it is equal to \eqref{eq:FDA}.
$S_\mathrm c$, finally, captures correlations between zero and thermal modes if $0 < T < T_c$.

\subsection{Application to Polaron Spectra}

For impurities in ultracold gases, an important observable is the spectral function, which
may be measured by rf spectroscopy:
A hyperfine state of the impurity, which is initially non-interacting with the bath, is
transferred by an rf pulse to an interacting state (injection) or vice versa (ejection)
\cite{Liu2020}.
The fraction of impurity atoms transferred is then measured as a function of the detuning
$\omega$ of the pulse's frequency compared to the bare transition without a bath.
The measured response is proportional to the spectral function $A(\omega)$, which is the
Fourier transform of the dynamical overlap $S(t)$:
\begin{align*}
S(t) &= \langle e^{it \fock H_i} e^{-it\fock H_f} \rangle \\
A(\omega) &= \frac{1}{\pi} \Re \int_0^{\infty} dt \, e^{it\omega} S(t)
\end{align*}
where $\fock H_i$ and $\fock H_f$ are the Hamiltonians corresponding to initial and transferred hyperfine states (thus one
of them with an interacting, the other one with a non-interacting impurity) and the density matrix is
determined by $\fock H_i$.

If the medium is non-interacting and the impurity of infinite mass, the Hamiltonians are one-particle
operators and the dynamical overlap can thus be computed with the FDA.
This has been done successfully for Fermions \cite{Knap2012, Cetina2016, Liu2020}, while for Bosons,
only a heuristic approach using only the factors $S_0 S_\mathrm t$ was available \cite{Schmidt2016, Schmidt2018}.
A different approach using Bogoliubov theory \cite{Dzsotjan2020}, however, correctly predicted the necessity of the
correlation factor and becomes equivalent to our approach in the case of ideal polarons.
Here, we have obtained this result from a general bosonic FDA and thus treat the Bose polaron in the same way as the Fermi polaron.

We consider the case where the impurity interacts via a contact potential:
\begin{equation*}
H_a = \frac{\vec p^2}{2m_B} + \VIB_a(\vec{x})
\end{equation*}
where $\vec p, \vec x$ are a bath particle's momentum and position operators, $m_B$ its mass and $\VIB_a$ the contact potential
of scattering length $a$, defined through
\begin{equation*}
\VIB_a(\vec{r}) \phi(\vec{r}) = \frac{4\pi a}{2m_B} \delta^3(\vec{r}) \frac{\partial}{\partial r} r \phi(\vec{r}).
\end{equation*}
For injection spectroscopy, we have $H_i = H_0 = p^2/2m_B$ and $H_f = H_a$, for ejection spectroscopy, the other way round.
The (generalized) s-wave eigenstates read
\begin{align*}
\psi^a_k(\vec{r}) &= \frac{\sin(kr) - ak\cos(kr)}{r\pi \sqrt{2(1 + a^2 k^2)}} & \text{ for } k > 0 \text{ (continuum)} \\
\psi^a_\bound(\vec{r}) &= \frac{\exp(-r / a)}{r\sqrt{2\pi a}} & \text{ if } a > 0 \text{ (bound state)}\\
\Phi^a_0(\vec{r}) &= \sqrt{n_0} \Bigl(1 - \frac{a}{r}\Bigr) & \text{ (condensate)}
\end{align*}
where $n_0$ is the condensate density, given by $n_0 = n (1 - (T/T_c)^{3/2})$ if $T\le T_c$ in terms of total bath density $n$ and
critical temperature $T_c$.
The normalization of the continuum modes is such that $\int_{\mathbb{R}^3} \psi^a_k \psi^a_q = \delta^1(k - q)$.

The energies are $E_k = k^2 / 2m_B$, $E_\bound = -a^{-2} / 2m_B$ and $E_0 = 0$.
In the case of the zero mode, we will also need the energy in finite volume $E_0 \simeq \frac{4\pi a}{2m_B\Omega}$ because it
doesn’t vanish when multiplied with a macroscopic number of occupying particles%
\footnote{
	Actually, this formula for $E_0$ is true only for Neumann boundary conditions.
	These seem the most appropiate to describe an ultracold gas in a box, which in reality has a healing length smaller than the system
	size and thus assumes a rather flat shape with center density $\approx N/\Omega$.
	But also for Dirichlet boundary conditions, the difference in energy to the impurity-free system obeys
	$N_0(E^{(a)}_0 - E_0^{(0)}) \approx \frac{4\pi a}{2m_B}n_0$ if $n_0$ is taken to be the local density at the box center in the
	impurity-free system \cite{Guenther2021}.
}.

\subsubsection{Overlap Operator}

To compute the determinants, we need the matrix elements of $M$ and of $M\ket{\Phi_0}\bra{\Phi_0}M$ as well as
$\bra{\Phi_0} M \ket{\Phi_0}$.
To have the operator $n_B(H_i)$ in diagonal form, we use the eigenbasis of $H_i$.
Since the contact potential acts only on s-wave states, also $M$ acts only on s-wave states.
We will find it more convenient to use $\tilde M = e^{-itH_i} M = e^{-itH_f} - e^{-itH_i}$ instead of $M$,
since it has symmetric integral kernel (this is because the $\psi^a_k$ were chosen real).

We proceed as in Ref.\ \cite{Drescher2021} and expand the eigenstates of the initial Hamiltonian in those of
the final one:
\begin{align*}
	\psi_k^i &= \alpha^{}_{k\bound} \psi_\bound^f + \int_{\mathbb R_+} dq \, \alpha^{}_{kq} \psi_q^f \\
	\psi_\bound^i &= \int_{\mathbb R_+} dk \, \alpha^{}_{\bound k} \psi_k^f \\
	\Phi_0^i &= \Phi_0^f + \alpha^{}_{0\bound} \psi_\bound^f + \int_{\mathbb R_+} dk \, \alpha^{}_{0k} \psi_k^f
\end{align*}
where terms with $\psi_\bound^f$ apply only to injection and $a > 0$ and $\psi_\bound^i$ only to ejection and $a>0$ in a
Fermionic bath.
The coefficients $\alpha$ are listed in appendix \ref{sec:overlap_operator}.

This expansion leads to formulas like the following: 
\begin{align*}
\braket{\psi^i_k | \tilde M | \psi^i_{k'}} 
={}& \int_{\mathbb R_+} dq \: \alpha_{kq} \alpha_{k'q} \Bigl(e^{-itq^2/2m_B} - e^{-itk^2/2m_B} \Bigr) \\
&+ \alpha_{k\bound} \alpha_{k'\bound} \Bigl( e^{ita^{-2}/2m_B} - e^{-itk^2/2m_B} \Bigr).
\end{align*}
The results of solving these integrals can be found in appendix \ref{sec:overlap_operator}.
These analytical expressions for the matrix elements lead to a significant speedup when computing the spectra.

\section{Results}

\begin{figure*}[h]
  \includegraphics[width=\textwidth]{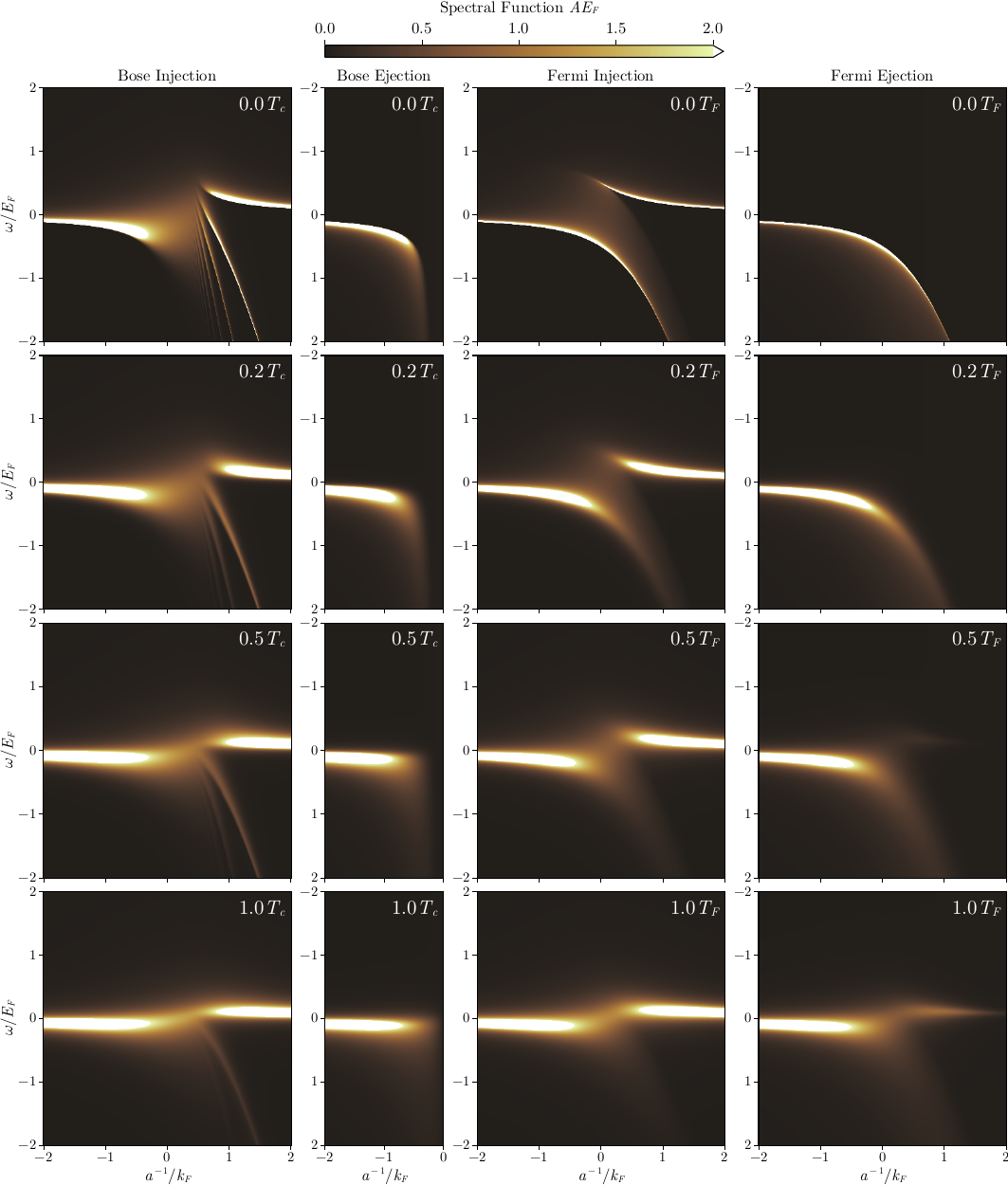}
  \caption{\label{fig:spectra_both}
	Spectral function of ideal polarons.
	For better comparison, the y axis of the ejection spectra is flipped.
	For Bose ejection, $a^{-1} > 0$ is impossible, since an ideal Bose gas
	would collapse onto the bound state.
	Note the different temperature units for Bosons and Fermions, so the temperature is not the same
	in the same row.
  }
\end{figure*}

With the explicit formulas for the overlap matrix elements, we can compute the dynamical overlap by discretizing
$k$-space,
and then obtain the spectra by Fourier transform.
The results are shown in Fig.\ \ref{fig:spectra_both}.
Also spectra for a spin-polarized Fermi gas obtained with the usual FDA are included here, both to facilitate direct comparison
and because the explicit formulas for $M$ allow us to compute them efficiently for more parameter values than
previously available.
Quantities are given in the units of a single-component Fermi gas, $k_F = (6\pi^2 n)^{1/3}$, $E_F = T_F = k_F^2/2m_B$, $t_F = 1/E_F$,
even for the Bose gas, except for its temperature, which is given in terms of the critical temperature $T_c \approx 0.44 T_F$.

\subsection{Zero temperature} \label{sec:zero_temperature}

At zero temperature, the formulas for $\braket{\Phi_0 | M | \Phi_0}$ in the appendix yield an explicit formula for the dynamical overlap.

The case of injection is discussed in detail in Ref.\ \cite{Drescher2021}:
The spectrum may be understood as emerging from discrete lines associated to zero mode and bound state, which are subsequently
broadened by the continuum. The discrete lines are located at the zero mode energy $4\pi a n / 2m_B$, plus multiples of the
binding energy in the case of $a > 0$.
The continuum leads to broadening but also to a shift, such that near the resonance, the peak position does not diverge (as the
zero mode energy would), but remains close to zero – this must be the case due to the sum rule $\int d\omega \, \omega A(\omega) = 0$.
For weak coupling, the spectrum is broadenend and shifted less and follows a Lévy distribution:
\begin{align*}
	A_\text{inj}(\omega) &\approx L\left( \omega; \omega_0 = \frac{4\pi an}{2m_B}, c = \frac{32\pi a^4n^2}{2m_B} \right)
\end{align*}
where
\begin{align*}
	L(\omega; \omega_0, c) = (\omega > \omega_0) \sqrt\frac{c}{2\pi} \frac{1}{(\omega - \omega_0)^{3/2}} e^{-\frac{c}{2(\omega - \omega_0)}}
\end{align*}
is the Lévy probability density.
This leads to peak position, peak height and full width at half maximum
\begin{align*}
	\omega_\text{max} &= \omega_0 + \frac{c}{3} = \omega_0 + \left(\frac{2}{3\pi}\right)^3 (ak_F)^4 E_F \\
	A(\omega_\text{max}) &\approx \frac{0.46}{c} = \frac{16.13}{(a k_F)^4 E_F} \\
	\text{FWHM} &\approx 0.900 c = 0.39 (ak_F)^4 E_F.
\end{align*}
In particular, both the offset $\omega_\text{max} - \omega_0$ between peak position and zero mode energy and
the peak width decrease as $a^4$ and thus much faster than $\omega_0 \sim a$ itself.

For ejection spectroscopy,
remarkably, the Lévy formula holds, with opposite sign for $\omega_0$, not only for weak
but for arbitrary coupling:
here, overlap and spectrum are given exactly by
\begin{align*}
	S_\text{ej}(t) &= \exp \left(it \frac{4\pi a n}{2m} - 8a^2 n \sqrt{\frac{\pi i t}{2m}} \right) \\
	A_\text{ej}(\omega) &= L\left( \omega; \omega_0 = -\frac{4\pi an}{2m}, c = \frac{32\pi a^4n^2}{2m} \right).
\end{align*}
Near the resonance, the peak position $\omega_\text{max}$ and width thus diverge as $a^4$.
For a real gas, a stabilizing effect of the Boson repulsion will eventually set in and prevent the divergence.

\subsection{Non-zero temperature}

\begin{figure*}
	\includegraphics[width=\linewidth]{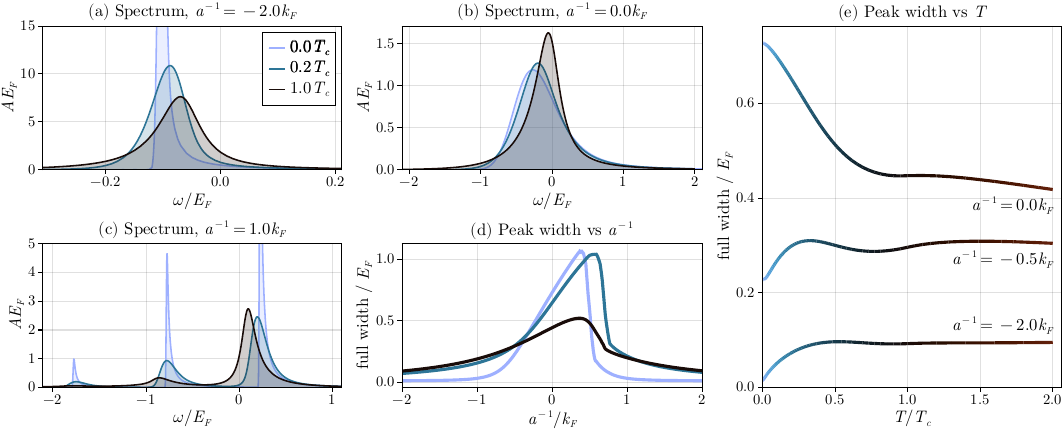}
	\caption{\label{fig:peak_widths}
		Peak shape of bosonic injection spectra.
		(a) Weak coupling. The sharp and asymmetric peak at absolute zero quickly broadens with temperature.
		At the same time, it shifts towards zero.
		(b) Strong coupling. Here, the situation is reversed: the broad zero temperature peak (note the different axis scale)
		becomes narrower for hotter systems.
		(c) Intermediate repulsive coupling. Here, the zero temperature peaks are sharp again and exhibit the
		hierarchy of different occupations of the bound state \cite{Drescher2021}.
		As temperature increases, more weight is transferred to the main peak, which corresponds to the repulsive polaron with no occupation of the bound state.
		(d) Full width near half maximum (we average over a region of 25--75\% of the maximum)
		of the main peak.
		The narrowing with increasing temperature observed in (b) takes place in a region around unitarity of size $\sim k_F$.
		(e) Peak width vs temperature. At weak coupling, most of the broadening takes place at low temperatures $T \lesssim 0.3 T_c$ and
		the width stays constant afterwards. At strong coupling, the width decreases until $\sim T_c$, then again at higher temperatures.
	}
\end{figure*}

Both for Fermions and Bosons, the sharp peaks at zero temperature and weak coupling quickly broaden
with temperature and are already significantly less sharp at $0.2 T_c$ or $0.2 T_F$, respectively.
For injection spectroscopy of Bosons near unitarity, however, the zero temperature peaks are very
broad and actually become sharper as temperature is increased.
This behaviour is shown in more detail in Fig.\ \ref{fig:peak_widths}.

The effect may be understood as follows:
The spectral function measures, essentially, the overlap of states initially present in the system
with the eigenstates of the final Hamiltonian.
At strong coupling, the low-energy eigenstates of non-interacting and interacting Hamiltonian are
very different, which leads to the broad peaks at low temperature.
As temperature increases, two effects compete:
On the one hand, the initial state becomes a mixture of many modes, and the spectrum can be considered a
combination of their respective spectra. This is the usual temperature broadening.
On the other hand, the high-energy states of the initial and final systems are not as different as
the low-energy states.
The spectrum may thus be understood as a combination of more, but narrower peaks as temperature increases.
This latter effect is thus found to be dominant near unitarity and for sub-critical temperatures.

Previous studies have not found the narrowing effect for different reasons.
In \cite{Dzsotjan2020}, spectra were evaluated at intermediate 
coupling $a^{-1} = -1 k_F$ and broadening with temperature was found, which is in agreement with Fig.\ \ref{fig:spectra_both}(d).
\cite{Guenther2018, Field2020} considered unitarity but employed a few-excitation approach, which predicts a branch splitting
with temperature that was argued to be an artifact of the approach in \cite{Field2020}.
It was suspected there that the amount of splitting might give an indication to the width of the actual peak, but this seems
questionable to us: if it were the case, the peaks should be split already at zero temperature to explain the broad unitarity peaks
found in \cite{Shchadilova2016}.

\subsection{Comparison with the Fermi polaron}

Comparing the Bosonic with the Fermionic injection spectra in Fig.\ \ref{fig:spectra_both}, the main differences are
the following:
\begin{enumerate}[(i)]
\item The Bosonic system allows for multiply occupied bound states, leading to a hierarchy of peaks on the repulsive side.
For a Fermionic bath, the Pauli principle allows only one bath atom to enter the bound state, and only two lines result.
\item As the resonance is crossed from attractive to repulsive coupling, the Bosonic spectrum becomes one
very broad peak which then dissects into the hierarchy of bound states.
The Fermionic spectrum remains a sharp peak, that eventually approaches the molecular binding energy.
\item For Fermions, the peak position $\omega_\text{max}$ at zero temperature is in excellent agreement with the ground-state
energy difference $\Delta E$ between final and initial system.
For Bosons, as demonstrated in section \ref{sec:zero_temperature},
$A(\omega)$ is non-zero only for $\omega > \Delta E$ and there is a shift between $\omega_\text{max}$
and $\Delta E$, which becomes negligible only for weak attractive coupling.
\end{enumerate}

Due to the Pauli blocking principle, the number of relevant bath excitations is much smaller for Fermions than for Bosons and
the former system can be well described by a variational wave function allowing for a single particle-hole excitation in the
Fermi sea \cite{Chevy2006}.
For Bosons, however, few-excitation approaches do not reproduce the observed broad peaks near unitarity and lead to spectra
more similar to the Fermionic ones \cite{Rath2013}. 

\subsection{Relevance of correlations}

The bosonic FDA consists of three factors, and while one can expect that close to zero temperature,
the zero mode factor is dominant, and close to critical temperature, the thermal factor is, it is
less clear, which of the other two provides the leading correction in both cases,
and if any of them can be neglected altogether in certain regimes.

\begin{figure}
	\includegraphics[width=\columnwidth]{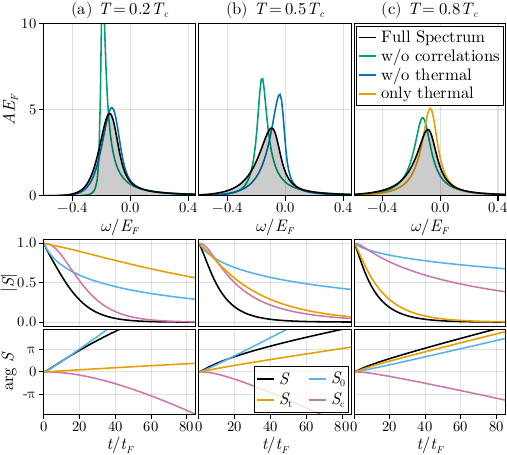}
	\caption{\label{fig:A0c_A0t}
		Role of the different terms in the bosonic FDA. For three temperatures, the injection spectrum $A(\omega)$ and overlap $S(t)$ at
		$a^{-1} = -1.0 k_F$ are shown together with curves resulting from taking into account only some of the three factors in
		eq.\ \eqref{eq:BFDA}.
		(a) At low temperature, the thermal factor has little relevance for the spectrum. Removing the correlation factor, 
		however, causes a drastic change.
		In the time evolution, the zero mode is important at short times, the correlation term at long times.
		(b) At intermediate temperature, removing either correlation or thermal term has a significant effect.
		Their absolute values are similar in the time evolution while the zero mode term is relevant at very short times.
		(c) Close to critical temperature, the thermal part alone constitutes a crude approximation.
		Adding the zero mode but no correlations does not constitute a significant improvement.
		It should be noted that different values of $a^{-1}$ can lead to quite different behaviour; e.g.\ the correlation term
		need not stay below 1 but may actually diverge at long times.
	}
\end{figure}

We cannot proceed in analogy to the discussion of \cite{Drescher2021} and regard the spectral function as
a folding of the Fourier transforms of the three factors, because they fail to be Fourier transformable
individually.
Also the product $S_\mathrm t S_\mathrm c$ fails to be so, hence the zero mode cannot be neglected without also neglecting the correlation term.
But the combinations $S_0 S_\mathrm c$ and $S_0 S_\mathrm t$ are transformable, so we can investigate what happens if
either the thermal or correlation factor is missing.
Fig.\ \ref{fig:A0c_A0t} shows the results for three temperatures.

At $0.2 T_c$, the correlation term is dominant:
leaving away the thermal term has little effect, while leaving away the
correlation term makes for a significant change.

At $0.5 T_c$, both deviate similarly and all terms are important.

At $0.8 T_c$, neglecting correlations causes an error of similar magnitude as
considering thermal states alone.

We conclude that zero mode and correlations must be taken into account in the entire
temperature range between $0$ and $T_c$, while the thermal factor becomes negligible near
zero temperature.

\section{Outlook}

We have formulated a general method for studying certain many-body observables of ideal
quantum systems undergoing Bose-Einstein condensation and applied it to compute spectral
functions of polarons in ultracold gases.
A simple derivation of Klich's formula in terms of Gaussian integrals suggests that it may be possible to find analogous
expressions for exponentials of other operators involving at most two creation and annihilation operators.
Such a generalization would provide a systematic way to study
many-body observables of the kind considered here for systems described by Bogoliubov-like Hamiltonians.
It would also
facilitate the application of coherent, squeezed and Gaussian state variational approaches to such systems.
In the context of ultracold polarons, these methods are widely employed, but are either restricted to zero temperature
(e.g.\ \cite{Shchadilova2016} and many later works)
or require additional approximations \cite{Dzsotjan2020}.

\acknowledgements

This work is funded by the DFG (German Research Foundation) under Project-ID 273811115 (SFB 1225 ISOQUANT)
and under Germany's Excellence Strategy EXC2181/1-390900948 (the Heidelberg STRUCTURES Excellence Cluster).

\appendix

\section{Derivation of the Bosonic FDA} \label{sec:derivation_bfda}

As mentioned in the main text, we use the following ensemble with a fixed number
of particles in the zero mode:
\begin{align*}
	\rho &= \frac{\delta_{\fock N_0, N_0} e^{-\beta \fock H_{gc}^\prime}}{Z},  \\
	Z &= \tr^\prime e^{-\beta \fock H_{gc}^\prime}
		= \detprime (1 - e^{-\beta H_{gc}^\prime})^{-1}.
\end{align*}

To be able to apply Klich’s formula, we need to bring the Kronecker delta to exponential
form by a Fourier transform,
\begin{equation*}
\delta_{\fock N_0, N_0} 
= \int_{-i \alpha}^{2\pi - i\alpha} \frac{d\nu}{2\pi} e^{i\nu (N_0 - \fock N_0)}
= \int_\gamma \frac{dz}{2\pi i} \frac{z^{N_0 - \fock N_0}}{z}
\end{equation*}
where $\alpha > 0$ does not affect the integral but ensures its exchangeability with the trace, and $\gamma$
is a circle around $0$ of radius $> 1$.

Applying Klich's formula, we obtain the following expression where the
single-particle operator corresponding to $\fock N_0$ 
is denoted $P_0$,
because it is the projector onto the ground state and
because the symbol $N_0$ is already used up for the number of condensed particles:
\begin{align*}
\langle e^{\fock A} e^{\fock B} \rangle 
&= \int_\gamma \frac{dz}{2\pi i}
	\frac{z^{N_0} \detprime (1 - e^{-\beta H_{gc}^\prime}) }{z \det (1 - z^{-P_0} e^{-\beta H_{gc}^\prime} e^A e^B)}.
\end{align*}
In the denominator, we use $z = \det z^{P_0}$ and $z^{P_0} = z P_0 + 1' = (z - 1) P_0 + 1$.
In the numerator, $\detprime (1 - X) = \det (1 - X')$:
\begin{align*}
\langle e^{\fock A} e^{\fock B} \rangle 
&= \int_\gamma \frac{dz}{2\pi i}
	\frac{z^{N_0} \det \bigl( 1 - \bigl(e^{-\beta H_{gc}^\prime}\bigr)'\bigr) }{\det \bigl( (z-1)P_0 + 1 - e^{-\beta H_{gc}^\prime} e^A e^B \bigr)}.
\end{align*}
Note that $z$ appears now only in the diagonal element associated to the zero mode.
With $e^{-\beta H_{gc}'} \bigl( 1 - \bigl( e^{-\beta H_{gc}'} \bigr)' \bigr)^{-1} = P_0 + n_B'$,
we arrive at
\begin{align*}
\langle e^{\fock A} e^{\fock B} \rangle 
&= \int_\gamma \frac{dz}{2\pi i}
	\frac{z^{N_0}}{\det ((z - 1) P_0 + X)}, \\
X &= 1' - (P_0 + n_B') (e^A e^B - 1).
\end{align*}

To separate $z$ from the determinant, we use the following formula for rank-one perturbations (see next appendix):
For any operator $A$ and scalar $\alpha$,
\begin{equation*}
\det (\alpha P_0 + A) = \alpha \detprime A + \det A.
\end{equation*}
Applying this formula takes out the $z$ from the determinant such that
the contour integral can be performed:
\begin{align*}
\langle e^{\fock A} e^{\fock B} \rangle 
&= \int_\gamma \frac{dz}{2\pi i}
	\frac{z^{N_0}}
	{(z - 1) \detprime X + \det X} \\
&= \frac{1}{\detprime X}
	\left(1 - \frac{\det X}{\detprime X}\right)^{N_0} \\
&\simeq \frac{1}{\detprime X}
	e^{-N_0 \frac{\det X}{\detprime X}}
\end{align*}
where we assumed the thermodynamic limit $N_0 \simeq \infty$ and used that
$\det X / \detprime X \sim 1/N_0$ if the thermodynamic limit is supposed to exist.

We now use one more determinant expansion formula to split off the zero mode (again, see next appendix):
For every $\alpha$,
\begin{equation*}
\det A = (\alpha + \bra{\phi_0} A \ket{\phi_0}) \detprime A - \alpha \detprime (A + \alpha^{-1} A \ket{\phi_0}\bra{\phi_0} A).
\end{equation*}

Applying this formula with $\alpha = N_0^{-1}$ yields
\begin{equation*}
e^{-N_0 \frac{\det X}{\detprime X}}
= e^{-N_0 \bra{\phi_0} X \ket{\phi_0} - 1 + \frac{\detprime (X + N_0 X \ket{\phi_0} \bra{\phi_0} X)}{\detprime X}}.
\end{equation*}

We have now removed the zero mode from all determinants and may take the thermodynamic limit:
We introduce the condensate wave function $\Phi_0 = \sqrt{N_0} \phi_0$,
which converges pointwise in position space in the thermodynamic limit.
Also, we drop the primes as the zero mode is no element of the Hilbert space in infinite volume.
We then arrive at the Bosonic FDA:
\begin{align*}
\langle e^{\fock A} e^{\fock B} \rangle &= S_0 S_\mathrm t S_\mathrm c \\
S_0 &= \exp \bra{\Phi_0} M \ket{\Phi_0} \\
S_\mathrm t &= \frac{1}{\det (1 - n_B M)} \\
S_\mathrm c &= \exp\left(\frac{\det (1 - n_B (M - M \ket{\Phi_0} \bra{\Phi_0} M))}{\det (1 - n_B M)} - 1 \right) \\
M &= e^A e^B - 1
\end{align*}
and likewise for more than two operators.
Here, expressions involving $\ket{\Phi_0}$, $\bra{\Phi_0}$ are defined via position space integrals.

\section{Determinant Expansion Formulas}

Here, the determinant expansion formulas used in appendix \ref{sec:derivation_bfda} to split off the zero mode are derived.
We first consider a finite-dimensional Hilbert space, then extend to infinite dimensions.

\subsection{Finite-dimensional case}
Consider a normalized vector $v$ and a linear map $A$, which we decompose as
\begin{equation*}
A = a_0 v v^\dagger + \phi v^\dagger + v \psi^\dagger + A' \\
\end{equation*}
with $\phi, \psi$ orthogonal to $v$ and $A'v = v^\dagger A' = 0$.
Then, for every scalar $\alpha$,
\begin{gather}
\det (\alpha v v^\dagger + A) = \det A + \alpha \detprime A \label{eq:det1} \\
\det A = (a_0 + \alpha) \detprime A - \alpha \detprime(\alpha^{-1} \phi \psi^\dagger + A) \label{eq:det2}
\end{gather}
where $\detprime$ denotes the determinant over the subspace orthogonal to $v$,
i.e.\ $\detprime (1 + X) = \det (1 + X^\prime)$.
The first formula splits off a rank-one projector, but keeps a $d$-dimensional determinant,
the second one reduces a $d$-dimensional determinant to two $d-1$-dimensional ones.

\paragraph{Notation}
For the proof, we make use of multivectors, i.e.\ elements of the Grassmann outer algebra of the vector space.
Multivectors of maximal grade $d$ (i.e.\ of the form $v_1 \wedge \dots \wedge v_d$) are called volume vectors.
Between linear maps $A_1, A_2, \dots$ on our vector space, we introduce the following symmetric, associative and distributive product:
\begin{equation*}
(A_1 \times \dots A_k) (v_1 \wedge \dots v_k) \defeq \frac{1}{k!} \sum_\sigma A_{\sigma(1)}(v_1) \wedge \dots A_{\sigma(k)}(v_k)
\end{equation*}
for a basis $k$-vector $v_1 \wedge \dots v_k$ and linearly extended to arbitrary vectors (the sum runs over permutations of $(1, ... k)$).
In particular,
$ A^{\times k} \defeq A \times \dots \times A $
is the natural action of $A$ on $k$-vectors,
\begin{equation*}
A^{\times k}(v_1 \wedge \dots v_k) = A(v_1) \wedge \dots A(v_k).
\end{equation*}
Note that this is zero if $k > \rank A$.
The determinant is the eigenvalue of the action on volume vectors:
\begin{equation*}
A^{\times d}(V) = \det(A) V
\end{equation*}
for every volume vector $V$.
The contraction of a vector and a multivector is
\begin{equation*}
a \cdot (v_1 \wedge \dots v_k) = \sum_i (-1)^{i-1} (a \cdot v_i) v_1 \wedge \dots \hat v_i \wedge \dots v_k
\end{equation*}
and linearly, the hat denoting omission.
The result is a $k-1$ vector orthogonal to $a$, i.e.\ one whose contraction with $a$ is zero.

\paragraph{Proof}
With this notation, one easily checks that for vectors $a, b$ and $B = A_2 \times \dots A_k$,
and any $k$-vector $V$,
\begin{equation*}
(ab^\dagger \times B)(V) = \frac{1}{k} a \wedge B(b \cdot V).
\end{equation*}
Combining this with the binomial theorem and the rank property yields
\begin{equation*}
(ab^\dagger + A)^{\times k}(V) = A^{\times k}(V) + a \wedge A^{\times k-1} (b \cdot V).
\end{equation*}

To derive the first determinant formula, we apply this with $v = a = b$ and $k=d$.
Let $d' = d-1$ and  $V' = v \cdot V$, such that $V = v \wedge V'$ with $V'$ orthogonal to $v$.
Then,
\begin{equation*}
(\alpha v v^\dagger + A)^{\times d}(V) 
= A^{\times d}(V) + \alpha v \wedge A^{\times d'}(V').
\end{equation*}
In the last expression,
$V'$ is a maximum grade vector on the subspace orthogonal to $v$,
so $A^{\times d'} (V') = \detprime(A) \, V'$ and we obtain \eqref{eq:det1}.

For the second formula,
we compute $\det A$ by inserting the expansion of $A$ in the definition of the
determinant:
\begin{align*}
A^{\times d}(V)
&= A(v) \wedge A^{\times d'}(V') \\
&= (a_0 v + \phi) \wedge (v \psi^\dagger + A')^{\times d'}(V') \\
&= (a_0 v + \phi) \wedge ({A'}^{\times d'}(V') + v \wedge {A'}^{\times d'-1}(\psi \cdot V') \\
&= a_0 v \wedge A'^{\times d'} (V') \\
&\phantom{=} + \phi \wedge v \wedge {A'}^{\times d'-1}(\psi \cdot V')\\
&= a_0 v \wedge A'^{\times d'} (V') \\
&\phantom{=} - v \wedge ((\phi \psi^\dagger + A')^{\times d'} - {A'}^{\times d'})(V')\\
&= (a_0 + 1) \detprime(A') V - \detprime(\phi \psi^\dagger + A') V,
\end{align*}
i.e.
\begin{align*}
\det A
&= (a_0 + 1) \detprime A - \detprime(\phi \psi^\dagger + A),
\end{align*}
By scaling $A$ (including $a_0$, $\phi$, $\psi$) with a factor $\alpha^{-1}$, this generalizes to \eqref{eq:det2}.

\subsection{Infinite-dimensional case}

For an infinite-dimensional space, the determinant is defined as Fredholm determinant (see e.g.\ \cite{Simon2005}).
Then, $\det (1 + T)$ is defined if and only if $T$ is trace class and this expression
is trace-norm-continuous.
For formulas \eqref{eq:det1}, \eqref{eq:det2},
every trace class operator $T$ can be approximated in the trace norm by a finite-rank
operator $N$.
If we have $A' = 1+N$, the determinants reduce to finite-dimensional ones and the formulas apply.
By continuity, they also apply to $A' = 1 + T$.

The determinants appearing in the main text are indeed well-defined:
As can be seen from the formulas in appendix \ref{sec:overlap_operator}, 
the integral kernels of $M$ and $M\ket{\Phi_0} \bra{\Phi_0} M$ are symmetric, continuous and increasing
at most polynomially as $k, k' \rightarrow \infty$.
Since $n_{B/F}$ are diagonal and decaying exponentially in the same basis,
$n^{1/2} M n^{1/2}$ and $n^{1/2} M \ket{\Phi_0} \bra{\Phi_0} M n^{1/2}$ have symmetric and continuous
integral kernels decaying sufficiently fast and thus are trace class.
Numerically, the determinants can be evaluated by approximation with finite-rank operators,
for instance by writing operators as integral operators and discretizing the integrals.

\section{Overlap Operator} \label{sec:overlap_operator}

For this appendix, we set $2m_B = n_0 = 1$.

The coefficients of expanding the eigenstates of the initial Hamiltonian in those of the final one are, for injection and
ejection respectively (c.f.\ \cite{Drescher2021}),

\begin{alignat*}{2}
	% \MoveEqLeft \textit{Injection:} & \MoveEqLeft \textit{Ejection:} \\
	\alpha^\text{inj}_{k,q} &= \frac{1}{\sqrt{1 + a^2 q^2}} \biggl( \delta(k-q) & {} + \mathscr{P} \frac{2akq}{\pi (q^2 - k^2)}\biggr)
	&= \alpha^\text{ej}_{q, k}  \\
	\alpha^\text{inj}_{k,\bound} &= \frac{2k a^{3/2}}{\sqrt{\pi} (1 + a^2 k^2)}
	&&= \alpha^\text{ej}_{\bound,k} \\
	\alpha^\text{inj}_{0, k} &= \mathscr{P} \frac{2a\sqrt{2}}{k \sqrt{1 + a^2 k^2}}
	\qquad; & - \mathscr{P} \frac{2a\sqrt{2}}{k} &= \alpha^\text{ej}_{0, k} \\
	\alpha^\text{inj}_{0,\bound} &= 2a \sqrt{2\pi a}
	&&
\end{alignat*}

The calculation of the matrix elements for the overlap operator $\tilde M$ proceeds as
in Ref.\ \cite{Drescher2021}, leading to the following results.
There, $\erfcx z = e^{z^2} (1 - \erf z)$ is the scaled complementary error function and
$\symm_{k \leftrightarrow k'} f(k, k') = f(k, k') + f(k', k)$.

\begin{widetext}

\subsubsection{Injection}

\begin{align*}
\braket{\psi^0_k | \tilde M | \psi^0_{k'}}
&= \frac{2k k'}{\pi} \Biggl[ 
	\frac{a^{-1}}{(k^2 + a^{-2})({k'}^2 + a^{-2})} \erfcx(-a^{-1} \sqrt{it})
	+ \symm_{k \leftrightarrow k'} \frac{1}{(k^2 - {k'}^2)(k^2 + a^{-2})} e^{-itk^2} \Bigl( ik\erf(ik\sqrt{it}) + a^{-1} \Bigr)
	\Biggr]
\\
\braket{\psi^0_k | \tilde M | \Phi^0_0}
&= \frac{2\sqrt{2} a}{1 + a^2 k^2} \Biggl[
	a^2 k \erfcx(-a^{-1} \sqrt{it}) 
	+ e^{-itk^2} \Bigl( ia\erf(ik\sqrt{it}) + \frac{1}{k} \Bigr) - \frac{1 + a^2 k^2}{k}
	\Biggr]
\\
\braket{\Phi^0_0 | \tilde M | \Phi^0_0}
&= -it 4\pi a + 4\pi a^3 \Bigl( \erfcx(-a^{-1}\sqrt{it}) - 1 - \frac{2}{\sqrt \pi} a^{-1} \sqrt{it} \Bigr).
\end{align*}
The last expression was already derived in \cite{Drescher2021} and yields the exact injection overlap for Bosons at zero temperature.

\subsubsection{Ejection}

\begin{align*}
\braket{\psi^a_q | \tilde M | \psi^a_{q'}}
&= \frac{2}{\pi} \frac{qq'}{\sqrt{q^2 + a^{-2}} \sqrt{{q'}^2 + a^{-2}}}
	\symm_{q \leftrightarrow q'} \frac{e^{-itq^2}}{q^2 - {q'}^2} \Bigl( iq\erf(iq\sqrt{it}) - a^{-1} \Bigr)
\\
\braket{\psi^a_b | \tilde M | \psi^a_q}
&= \frac{2 \sqrt{a^{-1}} q}{\sqrt \pi (a^{-2} + q^2)^{-3/2}} \Biggl[
	- a^{-1} \erfcx(a^{-1}\sqrt{it})
	+ e^{-itq^2} \Bigl( a^{-1} - iq\erf(iq\sqrt{it}) \Bigr)
\Biggr]
\\
\braket{\psi^a_b | \tilde M | \psi^a_b}
&= (1 + 2ita^{-2}) \erfcx(a^{-1} \sqrt{it}) - \frac{2a^{-1}}{\sqrt \pi} \sqrt{it} - e^{ita^{-2}}
\\
\braket{\Phi^a_0 | \tilde M | \psi^a_q}
&= \frac{2a\sqrt{2}}{q\sqrt{1 + a^2 q^2}} \Biggl[
	1 - e^{-itq^2} + iaq e^{-itq^2} \erf (iq\sqrt{it})
\Biggr]
\\
\braket{\Phi^a_0 | \tilde M | \Phi^a_0}
&= it 4\pi a - 8a^2 \sqrt{\pi i t}.
\end{align*}
$\braket{\Phi^a_0 | \tilde M | \psi^a_b}$ is not computed because it is irrelevant:
For ideal Bosons and $a > 0$, ejection spectroscopy does not make sense as the initial state would be collapsed.
\end{widetext}

\end{document}